\documentclass[final,1p,times]{elsarticle}

\usepackage{amssymb}
\usepackage{amsthm}
\usepackage{amsmath}
\usepackage{mathrsfs}
\usepackage{xcolor}
\usepackage{hyperref}
\usepackage{babel}
\usepackage{bm}
\usepackage{graphicx}
\usepackage{subcaption}

\usepackage{algorithm}  
\usepackage{algpseudocode}  
\usepackage{amsmath}

\newtheorem{theorem}{Theorem}
\newtheorem{proposition}[theorem]{Proposition}

\newtheorem{definition}{Definition}

\journal{}
\begin{document}

\begin{frontmatter}

\title{New Bayesian method for estimation of Value at Risk and Conditional Value at Risk}

\author{Jacinto Martín}\ead{jrmartin@unex.es}
\author{M. Isabel Parra}\ead{mipa@unex.es}
\author{Eva L.  Sanjuán}\ead{etlopez@unex.es}
\author{Mario M. Pizarro\corref{cor}}\ead{mariomp@unex.es}\cortext[cor]{Corresponding author}

\address{Department of Mathematics, University of Extremadura}
\date{Submitted to  Econometrics and Statistics}

\begin{abstract}

Value at Risk (VaR) and Conditional Value at Risk (CVaR) have become the most popular measures of market risk in Financial and Insurance fields.  However,  the estimation of both risk measures is challenging,  because it requires the knowledge of the tail of the distribution.  Therefore,  tools from Extreme Value Theory are usually employed,  considering that the tail data follow a Generalized Pareto distribution (GPD).  Using the existing relations from the parameters of the baseline distribution and the limit GPD's parameters,  we define highly informative priors that incorporate all the information available for the whole set of observations.  We show how to perform Metropolis-Hastings (MH) algorithm to estimate VaR and CVaR employing the highly informative priors, in the case of exponential,  stable and Gamma distributions.  Afterwards,  we perform a thorough simulation study to compare the accuracy and precision provided by three different methods.  Finally,  data from a real example is analyzed to show the practical application of the methods. 

\end{abstract}

\begin{keyword}

Value at Risk \sep Conditional Value at Risk  \sep generalized Pareto distribution \sep  Bayesian inference \sep Informative priors \sep  MH algorithm

\end{keyword}

\end{frontmatter}

\section{Introduction}\label{sec1}

Extreme value theory (EVT) is a branch of Statistics which comprehends the statistical tools that allow us to model and forecast events that are more extreme than any previously observed, called extreme values. It is widely employed in Climatology, to study extreme temperatures \cite{Garcia2021}, rainfall or floods \cite{Garcia2018}; but also in Finance and Insurance, mainly in Risk Management,  for estimation of financial reserves in insurance \cite{Longin1998, Chinhamu2015, Embrechts2013,Magnou2017}. 

A distribution that plays an important role in EVT is the Generalized Pareto distribution.  Due to its importance, several methods to estimate shape and scale parameters of the GPD have been proposed. Classical methods 
%like method of moments, method of weighted moments, maximum likelihood...a review of such methods can be found in \cite{ZeaBermudez2010, ZeaBermudez2010b}. However, all of them 
require conditions on the shape parameter for the asymptotic theory on which they are based, as it is shown in \cite{ZeaBermudez2010b}. That is why Bayesian inference is advisable. 

There have been some attempts to estimate the parameters of the GPD employing Bayesian techniques.  The first one corresponds to %\cite{Arnold1989}, who employed conjugate prior distributions,  
\cite{ZeaBermudez2003},  who employed Markov chain Monte Carlo (MCMC) methods,  \cite{Diebolt2005} estimated shape parameter when it is positive and \cite{Castellanos2007} applied Metropolis-Hastings algorithm and Jeffrey`s prior distribution. More recently, \cite{Martin2022} employed all the data available of the baseline distribution to perform estimations, based on existing relations between the parameters of the baseline distribution and the parameters of the limit GPD, applying the same strategy developed before in \cite{Martin2020} for block maxima and Gumbel distribution. 

%Besides, the problem of the choice of the threshold has also been studied by several authors.  %Reference \cite{Friggesi2002} proposed a method to estimate the threshold $u$ based on a mixture model for all data where the GPD is a component and a central model,  and also a dynamically weighted mixture model in which the central model is the Weibull distribution, \cite{Behrens2004} showed a truncated gamma model for non-extreme data and GPD for the tail data.  
%Reference \cite{Cabras2007} considered extreme observations to be outliers of a specified parametric model (a normal distribution),  and the threshold selection method is based on Bayesian outlier detection using partial posterior predictive p-value,  and \cite{Cabras2011} proposed an additive mixture model,  semi-parametric model for data below the threshold and GPD for the exceedances.  More recently,   \cite{Do2012} proposed a combined model with a nonparametric model for the central data and a GPD for the extreme data and \cite{Villa2017} offered a method with prior distributions for the threshold when data distribution is heavy-tailed.

In the field of Risk Theory, EVT is essential, because it allows us to study measures associated to those data that differ significantly from most observations, mainly to those belonging to the tail of the distribution. Most employed risk measures are Value at Risk (VaR) and Conditional Value at Risk (CVaR) \cite{Gilli2006, Bali2007,  Trzpiot2010, Park2016, Van2018,El2021, Kuang2022}. Some researchers have developed estimation strategies for VaR and CVaR, such as % \cite{Park2016}, who proposed and estimator based upon a non linear weighted least squares method, in  order to estimate VaR and CVaR for heavy tails; 
\cite{Korkmaz2018} proposed a three-parameter Pareto distribution to estimate VaR of a GPD, %\cite{El2021} employed GPD to compute VaR in a Generalized Autoregressive Score (GAS) model; 
\cite{Nolde2021} studied estimations for VaR and CVaR considering heavy tail distributions modelled by GPD; \cite{He2022} proposed a weighted random bootstrap method to estimate VaR by intervals, considering GPD. More recently, \cite{Faroni2022} showed the existing relation between VaR and CVaR for GPD and proposed a generalization of CVaR for high order risks %and \cite{Kuang2022} employed a generalized autoregressive conditional heteroskedasticity (GARCH) model, combined with a semiparametric approach, to forecast VaR for oil, using GPD, among others, as residual distribution. 

In this work, we will focus on estimating VaR and CVaR for different baseline distributions, employing Bayesian strategies similar to those proposed in  \cite{Martin2022} to estimate the parameters of the GPD, by using highly informative priors, built with all the information available in the whole dataset.

In Section \ref{sec2}, we introduce the definitions of VaR and CVaR and show their analytical expressions or how to compute them, when it is possible,  for GPD,  Gamma distribution,  and Stable distributions, in particular for Cauchy and Normal distributions.  In Section \ref{sec4}, we detail two Bayesian methods based on Metropolis-Hastings (MH) algorithm to estimate VaR and CVaR and propose a new method, considering highly informative priors, for exponential, stable and Gamma distributions. Afterwards, we perform a thorough simulation study to compare the accuracy provided by the three methods for the distributions considered, showing that the new method provides the best estimations for VaR and CVaR in Section \ref{sec5}. Finally, in Section \ref{sec6}, a real example is studied to provide a practical application of the methods.

\section{Risk measures}\label{sec2}

As we commented in the Introduction,  VaR and CVaR are usually the employed measures to describe the tail of a loss distribution in a financial context. % VaR represents the maximum loss expected over a certain period,  given a certain confidence level,  and  it is the standard measure of risk used by financial institutions \cite{Kourouma2011}. \cite{Artzner1999} exposed some theoretical deficiencies of VaR, and concluded that it is not a coherent measure because of the lack of subadditivity. Then, they proposed CVaR, as the expected size of a loss that exceeds VaR.  Therefore, precise estimations of such quantities are crucial.  

Consider a continuous random variable $X$ which represents the loss of an investment over a certain time horizon. 

\begin{definition}\label{def1}
 Given a parameter $0 < p < 1$, the Value at Risk ($\text{VaR}_p$) of $X$ is the $p$-quantile of the distribution $X$ 
\begin{equation}\label{eq1}
\text{VaR}_{p} (X)  = \inf \lbrace c : P \left( X \leq c \right) \geq p \rbrace
\end{equation} and
%\end{definition}
%\begin{definition}\label{def2}
%Given a parameter $0 < p < 1$, 
the Conditional Value at Risk ($\text{CVaR}_p$) of $X$ is 
\begin{equation}\label{eq2}
\text{CVaR}_p (X) = \text{E} \left[ X \mid X \geq \text{VaR}_p (X) \right] 
\end{equation}
\end{definition}

$\text{VaR}_p$ is the maximum loss expected in the $p\cdot 100\%$ of the best cases, while $\text{CVaR}_p$ is the expected loss conditioned to the loss is bigger than $\text{VaR}_p$. This measure is more sensitive to the shape of the tail of the loss distribution. 

\cite{McNeil2015} proposed an equivalent definition of $\text{CVaR}_p$
\begin{align}
\text{CVaR}_p (X) & = \dfrac{1}{1-p} \int_{p}^{1}{ \text{VaR}_q (X) dq} \nonumber  \\ 
& =  \dfrac{1}{1-p} \int_{\text{VaR}_p (X)}^{+\infty}{ xf(x) dx} \label{eq3}
\end{align}
where $f(x)$ is the density function of $X$. 

These measures depend on the distribution function of the observations, therefore in this work we show how to compute them for GPD and Gamma distribution. Afterwards, we focus on stable distributions and show how to compute VaR and CVaR for Cauchy and Normal distribution.

\subsection{Generalized Pareto distribution}\label{subsec1}

%As we mentioned before,  
Peaks-over-threshold (POT) method is based on the property that the distribution of data above a fixed threshold $u$ can be approximated by a properly scaled Generalized Pareto distribution, when $u$ tends to the endpoint of the distribution \cite{Balkema1974, Pickands1975}. Therefore, to estimate $\text{VaR}_p$, it is necessary that $p > u$. 

\begin{definition}\label{def3}
A random variable $X$ is distributed as a GPD with shape parameter $\xi$ and scale parameter $\sigma$ when its distribution function is 
\end{definition}
\begin{equation}\label{eq4}
\text{GPD} (\xi, \sigma; x ) = \left \lbrace \begin{matrix} 1 - \left( 1 + \xi \dfrac{x}{\sigma} \right)^{-1/\xi} & \text{if } \xi \neq 0 \\ &  \\ 1 - \exp \left( - \dfrac{x}{\sigma}  \right) & \text{if } \xi = 0   \end{matrix}   \right.
\end{equation}
where $\xi \in \mathbb{R},  \sigma > 0$, with support $x \geq 0 $ if $\xi \geq 0$ and $0 \leq x \leq - \dfrac{\sigma}{\xi}$ if $\xi < 0$.  

Notice that, when $\xi = 0$, GPD$(0,\sigma)$ is the Exponential distribution with parameter $\lambda = \dfrac{1}{\sigma}$. 

\begin{proposition}\label{prop1}
Let $X \sim \text{GPD} (\xi, \sigma)$ be a random variable. Given a parameter $0 < p < 1$, 
\begin{equation}\label{eq5}
\text{VaR}_p (X) = \dfrac{\sigma}{\xi} \left[ \left(1-p \right)^{-\xi} - 1 \right]
\end{equation}
\begin{equation}\label{eq6}
\text{CVaR}_p (X) = \dfrac{\sigma}{\xi} \left[ \dfrac{1}{1-\xi} \left( 1 - p \right)^{-\xi} - 1 \right]
\end{equation}

For $\xi=0$, $X \sim \text{Exp}(\lambda ) $ and
\begin{align}
\text{VaR}_p (X) & = - \dfrac{1}{\lambda} \ln \left( 1 - p \right) \label{eq7} \\
\text{CVaR}_p (X) & = \dfrac{1}{\lambda} \left[ 1 - \ln \left( 1 - p \right) \right] \label{eq8}
\end{align}
\end{proposition}

\subsection{Gamma distribution}\label{subsec2}

Gamma distribution is often employed to model financial data, so it is widely applied in the field of Risk Theory. 
%\begin{definition}\label{def4}
%A random variable $X$ is said to be distributed as a Gamma with shape parameter $\alpha > 0$ and scale parameter $\beta > 0$, when its density function is given by
%\begin{equation}\label{eq9}
%f(x) = \dfrac{\beta^\alpha}{\Gamma (\alpha)} x^{\alpha - 1}\exp \lbrace - \beta x \rbrace
%\end{equation} \end{definition} 
When $X \sim \Gamma (\alpha, \beta )$, it is impossible to find an analytic expression for $\text{VaR}_p$, because there is not such expression for the distribution function. However, by means of Equation \eqref{eq3}, it is possible to deduce $\text{CVaR}_p$ employing upper incomplete gamma function $\Gamma (a,x) = \int_{x}^{\infty}{t^{a-1}e^{-t}dt}$.  

\begin{proposition}\label{prop2}
Let $X \sim \Gamma (\alpha, \beta )$ be a random variable. Given a parameter $0 < p < 1$, the Conditional Value at Risk ($\text{CVaR}_p$) of $X$ is
\begin{equation}\label{eq10}
\text{CVaR}_p (X) = \dfrac{1}{\left( 1 - p \right) \Gamma \left( \alpha \right) \beta} \Gamma \left( \alpha + 1 ,  \beta \text{VaR}_p (X) \right)
\end{equation}
\end{proposition}

\subsection{Stable distributions}\label{subsec3}

Stable distributions are used to model physical and economic systems due to their important properties. Defined by \cite{Levy1925}, they have been studied and applied in a financial context \cite{ Embrechts2013,Nolan2020}.  Besides,  \cite{Yamai2002} showed different estimates for VaR and CVaR for stable distributions.  In previous work \cite{Martin2022},  we explained the relationship between stable distributions and GPD in order to compute estimations for the parameters of the GPD.

%\begin{definition}\label{def5}
%Let $Z$ be a random variable with parameters defined by its characteristic function:
%\begin{equation}\label{eq11} E[e^{itZ}]=\left\{\begin{array}{lr} exp\left\{-|t|^{\alpha}\left(1-i\beta \tan \dfrac{\pi\alpha}{2}sign(t)\right)\right\}, & \mbox{ if }\alpha \neq 1 \\ exp\left\{-|t|\left(1+i\beta \dfrac{2}{\pi}sign(t) \log|t|\right)\right\}, & \mbox{ if }\alpha=1 \end{array}\right.
%\end{equation} where the parameter $\alpha \in (0,2]$ is called the index of stability, and $\beta  \in [-1,1]$ is the skewness parameter. When $\beta=0$, the distribution is symmetric.
%A random variable $X$ is said to follow a stable distribution with parameters $a > 0$ and $b \in \mathbb{R}$ if it satisfies that \begin{equation}\label{eq12} 
%X = aZ+b. \end{equation} We denote $X \sim \text{Stable}(a,b)$.
%\end{definition}

%Notice that,  if $z_p$ is the $p$-quantile of $Z$,  then the $p$-quantile of $X$ can be computed as
%\begin{equation}\label{eq13}
%q_p = az_p + b
%\end{equation}

\begin{proposition}\label{prop3}
Let $Z \sim \text{Stable}(1,0)$ and $X = aZ + b$,  with parameters $a > 0$ and $b \in \mathbb{R}$.  
%Let $X \sim \text{Stable}(a,b)$ be a random variable. 
Given a parameter $0 < p < 1$,  
\begin{equation}\label{eq14}
\text{VaR}_p (X) = a \text{VaR}_p (Z) + b
\end{equation}
\begin{equation}\label{eq15}
\text{CVaR}_p (X) = a \text{CVaR}_p (Z) + b 
\end{equation}
\end{proposition}

\subsubsection{Normal distribution}\label{subsec31}

Normal distribution is one of the most used distributions in economic models and it has a great interest in finance and insurance.  Also,  this distribution is one of the best known stable distributions. % To compute the risk measures of any normal distribution,  we will first define $\text{VaR}_p$ and $\text{CVaR}_p$ for the standard Normal distribution.  

\begin{proposition}\label{prop4}
Let $X \sim \mathcal{N}(\mu , \sigma)$ be a random variable with mean $\mu \in \mathbb{R}$ and standard deviation $\sigma > 0$. Given a parameter $0 < p < 1$, %the Value at Risk ($\text{VaR}_p$) %is the $p$-quantile,
\begin{align}
\text{VaR}_{p}(X) & = \mu + \sigma z_p \label{eq18} \\
\text{CVaR}_p(X) & =  \mu +  \dfrac{\sigma}{(1-p)\sqrt{2\pi}}\exp \left \lbrace - \dfrac{1}{2}z_p^2 \right\rbrace \label{eq19}
\end{align} where $z_p$ is the $p$-quantile of $Z \sim \mathcal{N}(0,1)$.
%\begin{equation}\label{eq16}
%\text{VaR}_{p}(Z)  = z_p
%\end{equation}
%and the Conditional Value at Risk ($\text{CVaR}_p$) of $Z$ is
%\begin{align}\label{eq17}
%\text{CVaR}_{p}(Z) & = \dfrac{1}{(1-p)\sqrt{2\pi}}\exp \left \lbrace - \dfrac{1}{2}\text{VaR}_{p}(Z)^2 \right\rbrace
%\end{align}
\end{proposition}

%\begin{corollary}\label{coro5}
%Let $X \sim \mathcal{N}(\mu , \sigma)$ be a random variable with mean $\mu \in \mathbb{R}$ and standard deviation $\sigma > 0$. Given a parameter $0 < p < 1$, 
%\begin{align}
%\text{VaR}_{p}(X) & = \mu + \sigma \text{VaR}_p(Z) \label{eq18} \\
%\text{CVaR}_p(X) & = \mu + \sigma \text{CVaR}_p (Z)  \nonumber \\ &= \mu +  \dfrac{\sigma}{(1-p)\sqrt{2\pi}}\exp \left \lbrace - \dfrac{1}{2}\text{VaR}_{p}(Z)^2 \right\rbrace \label{eq19}
%\end{align}
%\end{corollary}

\subsubsection{Cauchy distribution}\label{subsec32}

Cauchy distribution is a stable distribution with heavy tails,  and it has the special characteristic that it does not present finite moments,  so it is not possible to determine its mean.  Then,  it is impossible to define the risk measure CVaR.  However,  unlike Normal distribution,  this time the risk measure VaR can be computed theoretically.  

%\begin{definition}\label{def6}
%A random variable $X$ is said to be distributed as a Cauchy with shape parameter $\delta > 0$ and location parameter $\gamma \in \mathbb{R}$, when its distribution function is given by 
%\begin{equation}\label{eq20}
%F(x) = \dfrac{1}{2} + \dfrac{1}{\pi} \text{atan} \left( \dfrac{x - \gamma}{\delta} \right)
%\end{equation} We denote $X \sim \mathcal{C}(\gamma,  \delta )$.
%\end{definition}

\begin{proposition}\label{prop6}
%Let $Z \sim \mathcal{C}(0,1)$ be a random variable.  
Let $X \sim \mathcal{C}(\gamma,  \delta )$ be a random variable.  Given a parameter $0 < p < 1$,  $\text{VaR}_p$ is computed by
\begin{equation}\label{eq22}
\text{VaR}_p (X)  = \gamma + \delta \text{tan} \left[ \pi \left( p - \dfrac{1}{2} \right) \right] 
\end{equation}
%\begin{equation}\label{eq21}
%\text{VaR}_p (Z) = \text{tan} \left[ \pi \left( p - \dfrac{1}{2} \right) \right] 
%\end{equation}
\end{proposition}

%\begin{corollary}\label{coro7}
%Let $X \sim \mathcal{C}(\gamma,  \delta )$ be a random variable. Given a parameter $0 < p < 1$, the Value at Risk ($\text{VaR}_p$) of $X$ is 
%\begin{equation}\label{eq22}
%\text{VaR}_p (X) = \gamma + \delta \text{VaR}_p (Z) = \gamma + \delta \text{tan} \left[ \pi \left( p - \dfrac{1}{2} \right) \right] 
%\end{equation}
%\end{corollary}

\section{Bayesian estimation methods}\label{sec4}

In this Section,  we will show how Bayesian methods can be applied to estimate VaR and CVaR for the GPD that models the tail of some distributions,  which we call baseline distributions.  We will focus on the cases when the baseline distribution is Exponential,  Gamma or Stable (Cauchy or Normal).  

Let $X$ be a random variable with distribution function $F(x; \theta)$ (baseline distribution) and define $u$ as a threshold value. The random variable $X_u = X-u \mid X > u$ represents the exceedances over $u$. We know \cite{Balkema1974,Pickands1975} that for a sufficiently high threshold $u$, the random variable $X_u$ follows a scaled Generalized Pareto distribution, GPD$(\xi, \sigma)$. 

Regarding to the threshold choice, in financial context,  the most usual way to select it is by fixing the probability below the threshold $p_u$,  that is,  by choosing the $p_u$-quantile of $X$. 

\begin{proposition}\label{prop8}
Let $u$ be the threshold defined as the $p_u$-quantile of $X$. For a $p$ parameter that satisfies $p > p_u$,  the VaR and CVaR of $X$ can be computed from the VaR and CVar of $X_u$, defined by Proposition \ref{prop1}. Therefore, 
\begin{equation}\label{eq23}
\text{VaR}_p \left( X \right)  =   u + \text{VaR}_{p_t} \left( X_u \right)
\end{equation}
\begin{equation}\label{eq24}
\text{CVaR}_p \left( X \right) =  u + \text{CVaR}_{p_t} \left( X_u \right) 
\end{equation} where $p_t = \dfrac{1-p}{1-p_u}$.
\end{proposition}

Usual Bayesian method to estimate a parameter is based on the Metropolis-Hastings (MH) algorithm.  We will consider non-informative prior distributions to estimate the parameters of the GPD using the random variable $X_u$ (details of the method can be consulted in \cite{Martin2022}) and apply Proposition \ref{prop8} to estimate VaR and CVaR.  However,  since this method only employs the observations above the threshold,  which are usually scarce,  it provides skewed estimates oftenly. 

To seize all the dataset,  the idea of using all the observations from the baseline distribution has already been employed in \cite{Martin2022,  Martin2020}.  We called this method Baseline MH method (BMH),  and in this work we apply it to the estimation of VaR and CVaR,  which can be computed straightforward.  In this case,  the problem can arise when we do not know the baseline distribution of the data,  which is quite usual in real situations.  

The new method we proposed tries to solve these difficulties.  It is also based on MH,  but employs can informative prior distribution for the parameters of the GPD,  obtained from the relationship between the parameters $\xi$ and $\sigma$ of the GPD and the parameters of the baseline distribution.  We called this method Informative Prior Baseline MH method (IPBMH) and it allows to compute VaR and CVaR for $X_u$ and by Proposition \ref{prop8},  the estimates for risk measures of $X$ can also be computed.  

In this method,  we give more weight to the tail values than to the values under the threshold.  In a first step,  BMH is used to estimate the parameters of the baseline distribution that will be used in the existing relationships.  With them, highly informative prior distributions are constructed for the parameters of the GPD, using only the excesses of the threshold. 

Next, this method is presented for each baseline distribution, since it is necessary to know the relationships between the baseline parameters and the GPD parameters.

Let $\mathbf{x}_u = \left( x_u^1 , ..., x_u^m \right)$ be a sample of size $m$ of the random variable $X_u$.

\subsection{Exponential distribution}

When the baseline distribution is an Exponential distribution with parameter $\lambda$,  the tail distribution $X_u$ is the same as the underlying distribution.  Therefore,  we can establish the relationship $\lambda = \dfrac{1}{\sigma}$.  

As a consequence,  the following prior distribution can be considered to estimate $\sigma$ $$\sigma \sim \mathcal{N}\left( \mu_\sigma, \nu_{\sigma} \right)$$ with $\mu_\sigma = \dfrac{1}{\lambda}$, where $\lambda$ will be estimated by BMH. 
%The posterior distribution is 
%
%\begin{align}\label{eq25}
%\pi \left( \sigma \mid \mathbf{x}_u \right) \propto \mathcal{L}\left( \sigma ; \mathbf{x}_u \right) \pi \left( \sigma \right) \propto \sigma^{-m}\exp \left \lbrace -\dfrac{1}{\sigma} \sum_{i=1}^m x_u^i - \dfrac{1}{2\nu_\sigma^2}  \left( \sigma - \mu_\sigma \right)^2 \right \rbrace
%\end{align}

As the prior distribution is highly informative, the probability of acceptance of the MH algorithm is only affected by the likelihood functions, so the ratio can be computed as 

\begin{align}\label{eq26}
r_\sigma & =  \left( \dfrac{\sigma^{(j)}}{\sigma^{\ast}} \right)^m \exp \left\lbrace \left(  \dfrac{1}{\sigma^{(j)}} - \dfrac{1}{\sigma^{\ast}}  \right) \sum_{i=1}^m x_u^i \right \rbrace 
\end{align}

\subsection{Stable distributions}

Let $Z\sim \text{Stable}(1,0)$ and $X = aZ + b$.  For a threshold $u$, it is satisfied that $Z_u \sim \text{GPD}(\xi_Z, \sigma_Z )$, and consequently $X_u \sim \text{GPD}(\xi_Z, a\sigma_Z )$. Therefore, it is enough to find estimates for the parameters $\xi_Z$ and $\sigma_Z$.  In \cite{Martin2022}, a simulation study is shown for the estimates of standard stable distributions Normal and Cauchy, for a threshold $u$ defined as the $p_u$-quantile.

\begin{table}[htp]
\begin{center}
\caption{Estimates for GPD's parameters}\label{tab1}%
\begin{tabular}{@{}lll@{}}
\hline
distribution & $\xi_Z $ & $\sigma_Z$  \\
\hline
$\mathcal{N}(0,1)$ & $-0.7+0.61p_u$ & $0.34+3.18(1-p_u) - 12.4(1-p_u)^2$ \\ 
$\mathcal{C}(0,1)$ & $1$ & $\dfrac{1}{\pi}(1-p_u)^{-1} $ \\ 
\end{tabular}
\end{center}
\end{table}

Table \ref{tab1} shows the estimates found. Then,  to estimate the GPD parameters, $\xi, \sigma$,  we can take the prior distributions
\begin{align*}
\xi \sim \mathcal{N} ( \xi_Z,  b_1 ),     \sigma \sim \mathcal{N} ( a \cdot \sigma_Z,  b_2 )
\end{align*} where $a$ will be estimated by BMH. In particular, $a$ is the scale parameter of stable distributions, so for the normal distribution it is equal to $\sigma$ and for the Cauchy distribution it is $\delta$. The coefficients $b_1$ and $b_2$ are shown in Table \ref{tab2}.

\begin{table}[htp]
\begin{center}
\caption{Coefficients of the prior distributions' hyperparameters}\label{tab2}%
\begin{tabular}{@{}lllll@{}}
\hline
\multicolumn{2}{l}{} & \multicolumn{3}{l}{$b_2 = \exp \lbrace c_1p_u^2 + c_2p_u + c_3 \rbrace $}  \\ 
\hline
distribution & $b_1$ & $c_1$ & $c_2$ & $c_3$ \\
\hline
$\mathcal{N}(0,1)$ & $0.03$ &  $-46.24$ & $83.55$ & $-41.58$ \\
$\mathcal{C}(0,1)$ & $0.065$ &  $323.57$ & $-588.51$ & $266.13$ \\
\end{tabular}
\end{center}
\end{table}

%The joint posterior distribution is
%\begin{align}\label{eq27}
% \pi ( \xi, \sigma \mid \mathbf{x}_u ) & \propto \sigma^{-m}\exp \left \lbrace -\frac{1}{2b_1^2} \left( \xi -\xi_Z \right)^2 -\frac{1}{2b_2^2} \left( \sigma - a \cdot  \sigma_Z \right)^2 \right \rbrace \prod_{i=1}^{m}\left( 1 + \xi \frac{x^i_u}{\sigma} \right)^{-(1+1/\xi)}
%\end{align}
%
%and marginal posterior distributions are
%\begin{align}\label{eq28}
% \pi ( \xi \mid \sigma, \mathbf{x}_u ) & \propto \exp \left \lbrace -\frac{1}{2b_1^2} \left( \xi - \xi_Z \right)^2  \right \rbrace \prod_{i=1}^{m}\left( 1 + \xi \frac{x^i_u}{\sigma} \right)^{-(1+1/\xi)} \nonumber \\ \\
% \pi ( \sigma \mid \xi, \mathbf{x}_u ) & \propto \sigma^{-m}\exp \left \lbrace  -\frac{1}{2b_2^2} \left( \sigma - a \cdot  \sigma_Z \right)^2 \right \rbrace \prod_{i=1}^{m}\left( 1 + \xi \frac{x^i_u}{\sigma} \right)^{-(1+1/\xi)} \nonumber
%\end{align}
%Finally, 
The ratios of MH algorithm for each parameter are
\begin{align}\label{eq29}
r_\xi &= \exp \left \lbrace \frac{1}{2b_1^2} \left( \left( \xi^{(j)}-\xi_Z \right)^2 -\left( \xi^{\ast}-\xi_Z \right)^2  \right) \right. \nonumber \\ & \left.- \left( 1 + \frac{1}{\xi^{\ast}} \right)\sum_{i=1}^m \ln \left( 1 + \xi^{\ast} \frac{x^i_u}{\sigma^{(j)}} \right) + \left( 1 + \frac{1}{\xi^{(j)}} \right)\sum_{i=1}^m \ln \left( 1 + \xi^{(j)} \frac{x^i_u}{\sigma^{(j)}} \right) \right \rbrace
\end{align}

\begin{align}\label{eq30}
r_\sigma &= \left( \frac{\sigma^{(j)}}{\sigma^\ast} \right)^m \exp \left \lbrace \frac{1}{2b_2^2} \left( \left( \sigma^{(j)}- a \cdot \sigma_Z \right)^2 -\left( \sigma^{\ast}-a \cdot  \sigma_Z \right)^2  \right)\right. \nonumber \\ &  \left. - \left( 1 + \frac{1}{\xi^{(j)}} \right)\sum_{i=1}^m \ln \left( 1 + \xi^{(j)} \frac{x^i_u}{\sigma^{\ast}} \right) + \left( 1 + \frac{1}{\xi^{(j)}} \right)\sum_{i=1}^m \ln \left( 1 + \xi^{(j)} \frac{x^i_u}{\sigma^{(j)}} \right) \right \rbrace
\end{align}

\subsection{Gamma distribution}

Let $X \sim \Gamma ( \alpha , \beta )$ be a random variable.  For a threshold $u$, the random variable of excesses $X_u$ follows a $\text{GPD}(\xi, \sigma)$.  A simulation study has been carried out, similar to that of stable distributions, with the aim of obtaining empirical relationships between $\alpha, \beta$ parameters and the GPD parameters.  Obtained relationships are
\begin{equation} \label{eq31}
\xi  = -0.032 + 0.014\alpha^{-1},   \qquad  \sigma  = \dfrac{1}{2}\beta^{-1}\left( 1 + \sqrt{\alpha} \right)
\end{equation} 
%\begin{equation}\label{eq32}
%\sigma  = \dfrac{1}{2}\beta^{-1}\left( 1 + \sqrt{\alpha} \right)
%\end{equation}

These empirical relationships allow the construction of highly informative prior distributions for $\xi$ and $\sigma$. Therefore, the prior distributions are
\begin{align*}
\xi \sim \mathcal{N} \left( \mu_\xi , \nu_\xi \right), \quad \sigma \sim \mathcal{N} \left( \mu_\sigma , \nu_\sigma \right)
\end{align*} where 
\begin{equation}\label{eq33}
\mu_\xi = -0.032+0.014\alpha^{-1}, \quad \mu_\sigma = \dfrac{1}{2}\beta^{-1}\left( 1 + \sqrt{\alpha} \right)
\end{equation}

%The joint posterior distribution is
%\begin{align}\label{eq34}
% \pi ( \xi, \sigma \mid \mathbf{x}_u ) & \propto \sigma^{-m}\exp \left \lbrace -\frac{1}{2\nu_\xi^2} \left( \xi -\mu_\xi \right)^2 -\frac{1}{2\nu_\sigma^2} \left( \sigma - \mu_\sigma \right)^2 \right \rbrace \prod_{i=1}^{m}\left( 1 + \xi \frac{x^i_u}{\sigma} \right)^{-(1+1/\xi)}
%\end{align}
%
%and marginal distributions are
%\begin{align}\label{eq35}
% \pi ( \xi \mid \sigma, \mathbf{x}_u ) & \propto \exp \left \lbrace -\frac{1}{2\nu_\xi^2} \left( \xi - \mu_\xi \right)^2  \right \rbrace \prod_{i=1}^{m}\left( 1 + \xi \frac{x^i_u}{\sigma} \right)^{-(1+1/\xi)} \nonumber \\ \\
% \pi ( \sigma \mid \xi, \mathbf{x}_u ) & \propto \sigma^{-m}\exp \left \lbrace  -\frac{1}{2\nu_\sigma^2} \left( \sigma - \mu_\sigma \right)^2 \right \rbrace \prod_{i=1}^{m}\left( 1 + \xi \frac{x^i_u}{\sigma} \right)^{-(1+1/\xi)} \nonumber
%\end{align}
In this case, the probability of acceptance is only affected by the likelihood functions, so the ratio can be computed as
\begin{align}\label{eq36}
r_{\xi, \sigma} &= \left( \dfrac{\sigma^{(j)}}{\sigma^{\ast}} \right)^m \cdot \prod_{i = 1}^{m} \left[ \left( 1 + \xi^{\ast} \dfrac{x_u^i}{\sigma^{\ast}} \right)^{-(1+1/\xi^\ast)} \cdot \left( 1 + \xi^{(j)} \dfrac{x_u^i}{\sigma^{(j)}} \right)^{1+1/\xi^{(j)}} \right]
\end{align}

\section{Results}\label{sec5}

In this Section,  we show the results of a thorough simulation study in order to compare the quality of the estimates for VaR and CVaR computed with the three Bayesian methods explained in previous section. 

We fixed the threshold as the quantile of $0.9$ order.  Therefore,  $p_u = 0.9$,  and we computed $\text{VaR}_{0.95}\left( X \right)$, $\text{CVaR}_{0.95} \left( X \right) $,  being $X$ the baseline distribution.  We developed the cases studied in Section \ref{sec4},  Exponential,  Stable (Cauchy and Normal) and Gamma distributions,  for different sizes and parameters. 

For each baseline distribution and fixed parameters,  samples of sizes $n = 2^i, i = 5,...,10$ were simulated. We obtained a Monte Carlo Markov chain (MCMC) of length 10000,  taking 3000 training values and 50 thinning values.  Finally,  each sequence was repeated 100 times,  and we computed the mean to estimate VaR or CVaR.

Next,  we provide graphs with the estimates of both risk measures computed by the three methods: MH,  BMH  and IPBMH,  in order to compare their accuracy. 

\subsection{Exponential distribution}

\begin{figure}[htp]
\begin{center}
\includegraphics[width=0.85\textwidth]{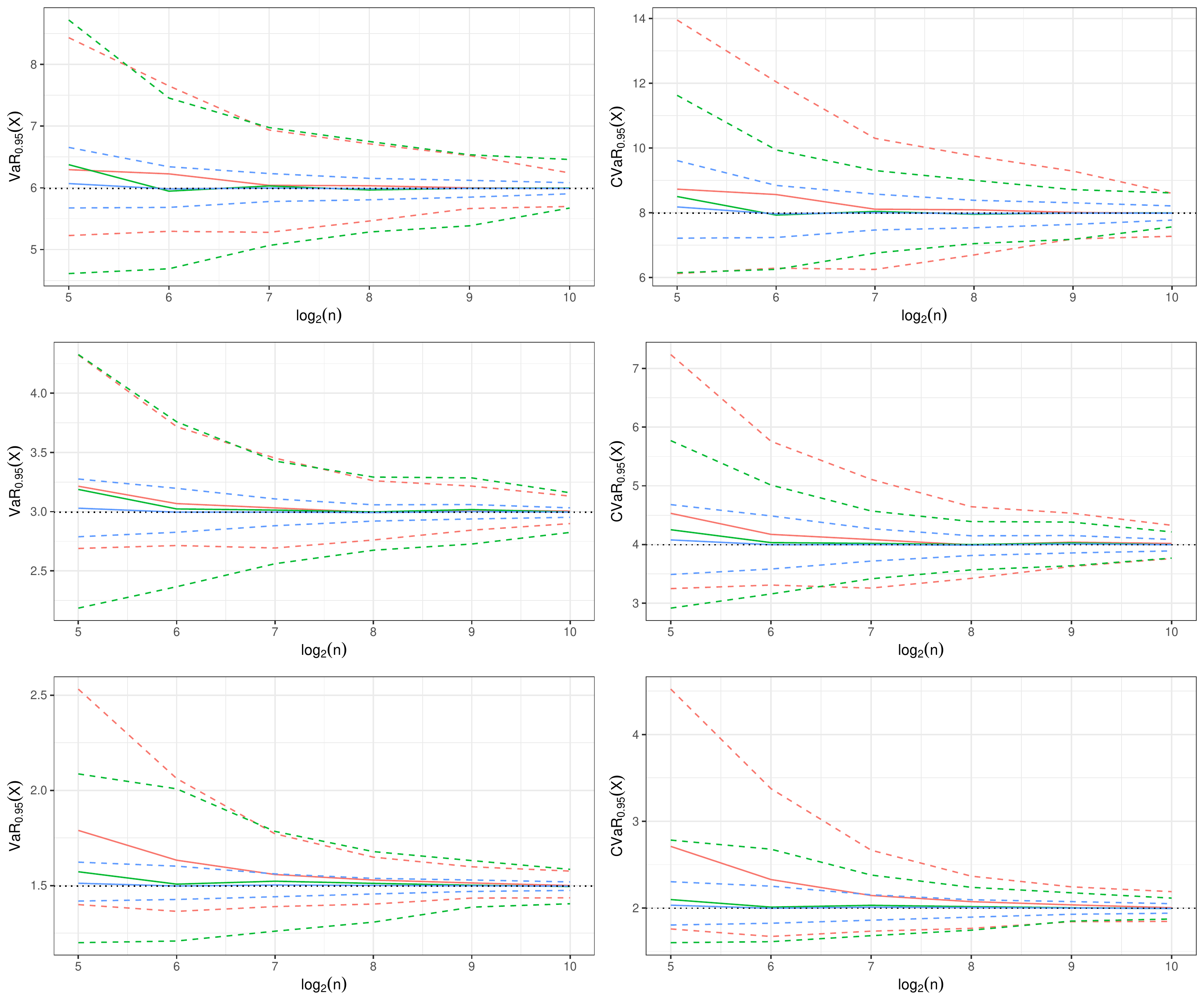}
\end{center}
\caption{Estimates of $\text{VaR}_{0.95} (X)$ (left charts) and $\text{CVaR}_{0.95} (X)$ (right charts), for $X \sim \text{Exp}(\lambda)$ with $\lambda =$ 0.5 (upper charts),  $\lambda =1$ (middle charts) and $\lambda =2$ (lower charts),  with 2.5\% and 97.5\% confidence bounds, by MH (red lines), BMH (green lines) and IPBMH (blue lines). Black lines represent the real value of each risk measure.}\label{fig1}
\end{figure}

Let $X\sim \text{Exp}(\lambda)$,  we considered $\lambda = 2^j, j = -2,-1,0,1,2$.  Figure \ref{fig1} shows the estimates for $\text{VaR}_{0.95}(X)$ (left) and $\text{CVaR}_{0.95}(X)$ (right),  along with 2.5\% and 97.5\% confidence bounds,  computed with MH (red lines),  BMH (green lines) and IPBMH (blue lines) for $\lambda = 0.5$ (upper charts),  $\lambda = 1$ (middle charts) and $\lambda = 2$ (lower charts).  Besides,  the real values of each risk measures are drawn as a black line.

As we can see,  MH is the method that offers the worst estimates,  especially when data are scarce,  while IPBMH offers the most accurate and precise estimates for both risk measures for any size of the dataset $n$.  

We also observed that high values of $\lambda$ produced more deviations in estimations form MH and BMH methods when $n$ is low.  Again,  estimates from IPBMH remain stable. 

\subsection{Stable distributions}

Let $X \sim \text{Stable}(a,b)$.  For stable distributions,  the location parameter $b$ does not have an influence on the estimates of the parameters of the GPD,  therefore it does not affect the estimates of the risk measures.  So,  we fixed $b = 0$.  In particular,  $\mu = 0$ and $\gamma = 0$ for Normal and Cauchy distribution, respectively.  

\begin{figure}[htp]
\begin{center}
\includegraphics[width=0.85\textwidth]{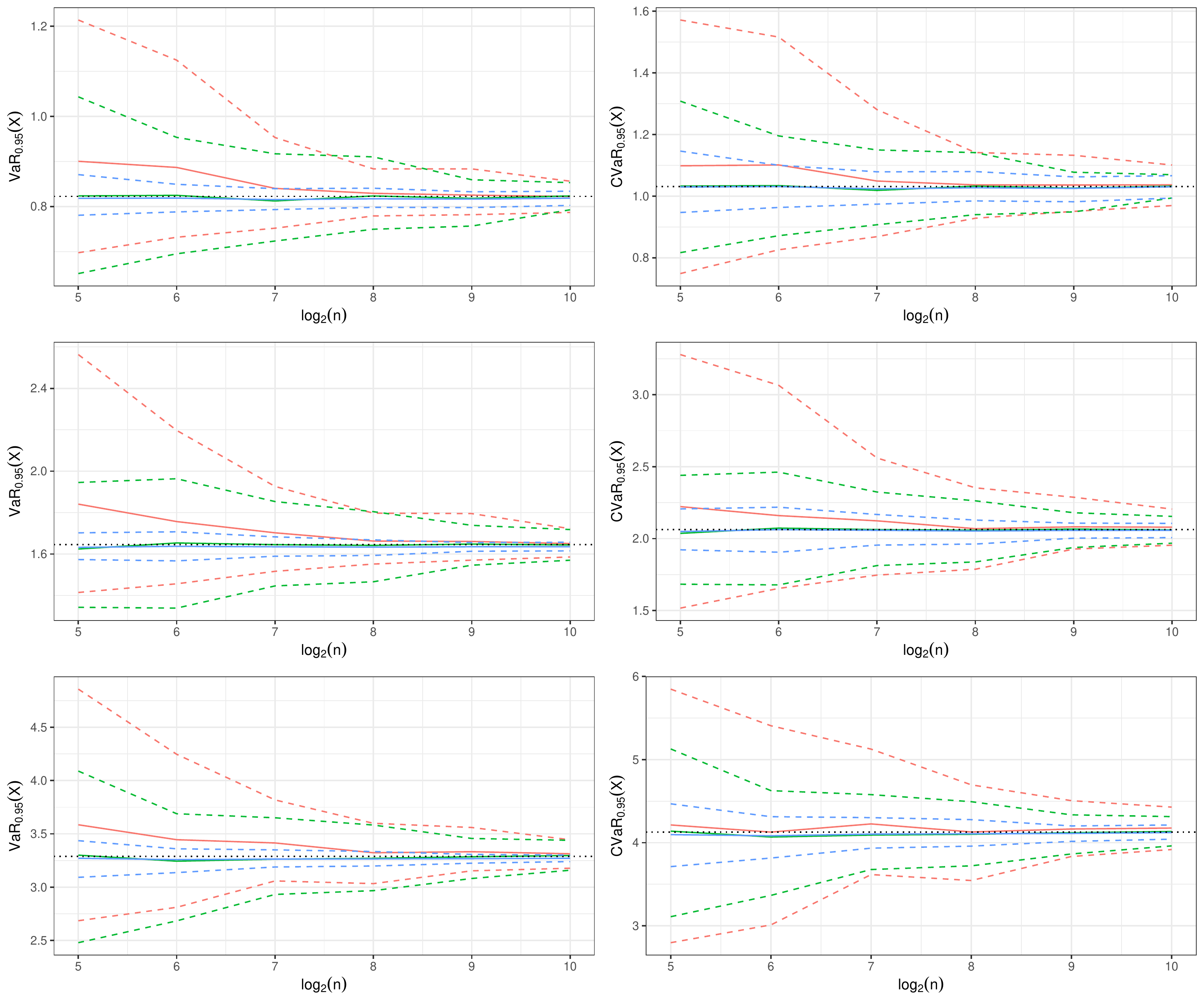}
\end{center}
\caption{Estimates of $\text{VaR}_{0.95} (X)$ (left charts) and $\text{CVaR}_{0.95} (X)$ (right charts), for $X \sim \mathcal{N}(0,\sigma)$ with $\sigma = 0.5$ (upper charts),  $\sigma = 1$ (middle charts) and $\sigma = 2$ (lower charts),  with 2.5\% and 97.5\% confidence bounds, by MH (red lines), BMH (green lines) and IPBMH (blue lines). Black lines represent the real value of each risk measure.}\label{fig2}
\end{figure}

\begin{figure}[htp]
\begin{center}
\includegraphics[width=\textwidth]{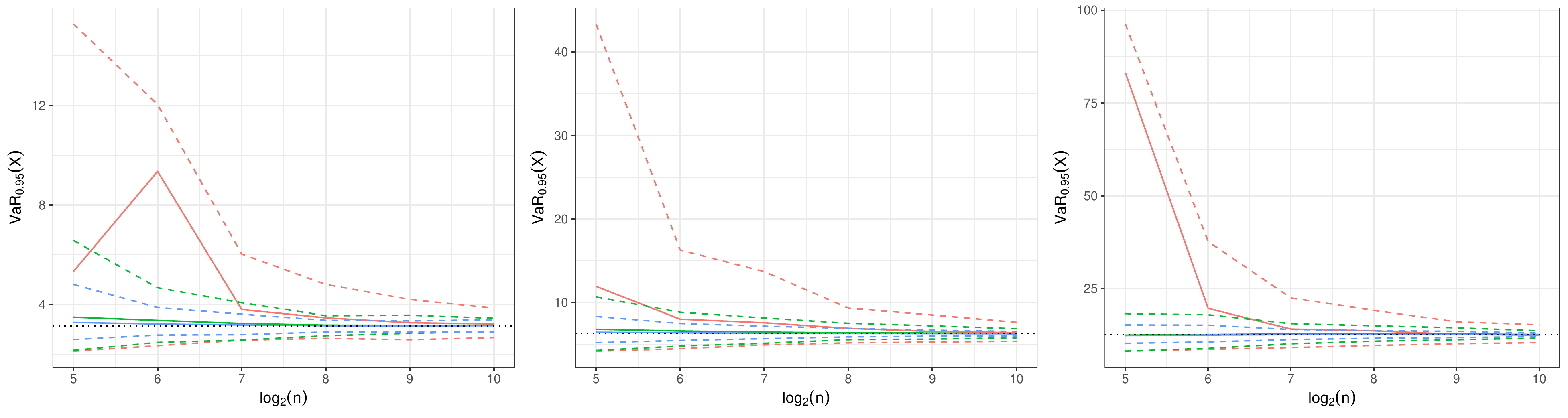}
\end{center}
\caption{Estimates of $\text{VaR}_{0.95} (X)$ for $X \sim \mathcal{C}(0,\delta)$ with $\delta = 0.5$ (left chart),  $\delta = 1$ (center chart) and $\delta = 2$ (right chart) , with 2.5\% and 97.5\% confidence bounds, by MH (red lines), BMH (green lines) and IPBMH (blue lines). Black lines represent the real value of each risk measure.}\label{fig3}
\end{figure}

The scale parameter $a$ does affect the estimates,  therefore we took $a = 2^j,  j = -2,-1,0,1,2$ ($\sigma$ for Normal,  $\delta$ for Cauchy).  The most representative cases $a = 0.5,1,2$ are shown in Figures \ref{fig2}-\ref{fig3}.  In particular,  Figure \ref{fig2} shows the estimates of VaR and CVaR for Normal distribution and Figure \ref{fig3} shows the estimates for VaR.  In both figures we can see how BMH and IPBMH provide very similar estimates for any size $n$,  for any value of the parameter.  The estimates are also quite accurate,  but IPBMH shows less dispersion than BMH.  In contrast,  MH provides the worst estimates,  especially when there is a lack of data.  

In particular, for the Cauchy distribution,  when the scale parameter $\delta$ is high and the size $n$ is low,  MH offers estimates of VaR with really high deviations from the real value.  

\subsection{Gamma distribution}

\begin{figure}[htp]
\begin{center}
\includegraphics[width=0.85\textwidth]{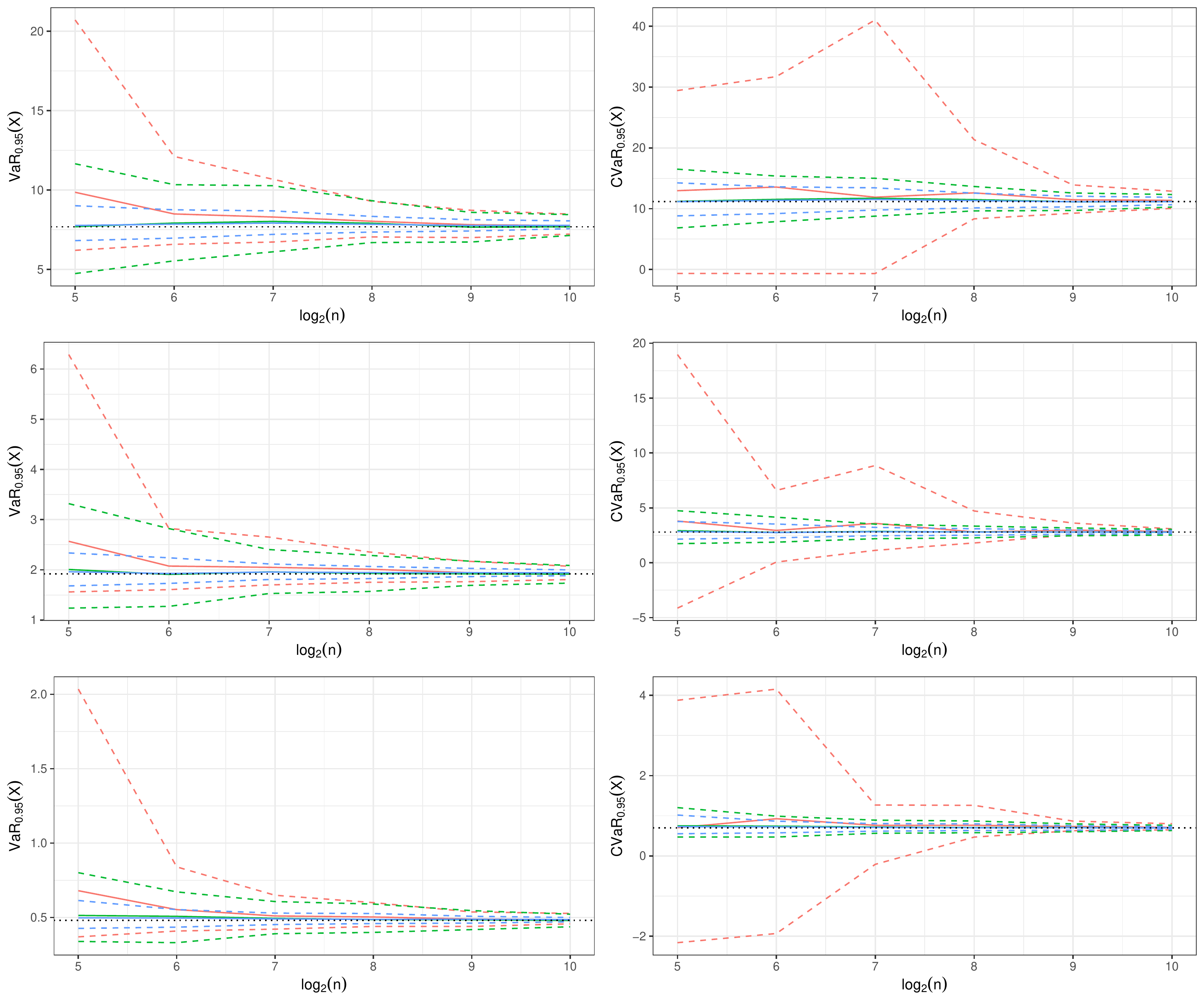}
\end{center}
\caption{Estimates of $\text{VaR}_{0.95} (X)$ (left charts) and $\text{CVaR}_{0.95} (X)$ (right charts), for $X \sim \Gamma(0.5,\beta)$ with $\beta = 0.25$ (upper charts),  $\beta = 1$ (middle charts) and $\beta = 4$ (lower charts),  , with 2.5\% and 97.5\% confidence bounds, by MH (red lines), BMH (green lines) and IPBMH (blue lines). Black lines represent the real value of each risk measure.}\label{fig4}
\end{figure}

\begin{figure}[htp]
\begin{center}
\includegraphics[width=0.85\textwidth]{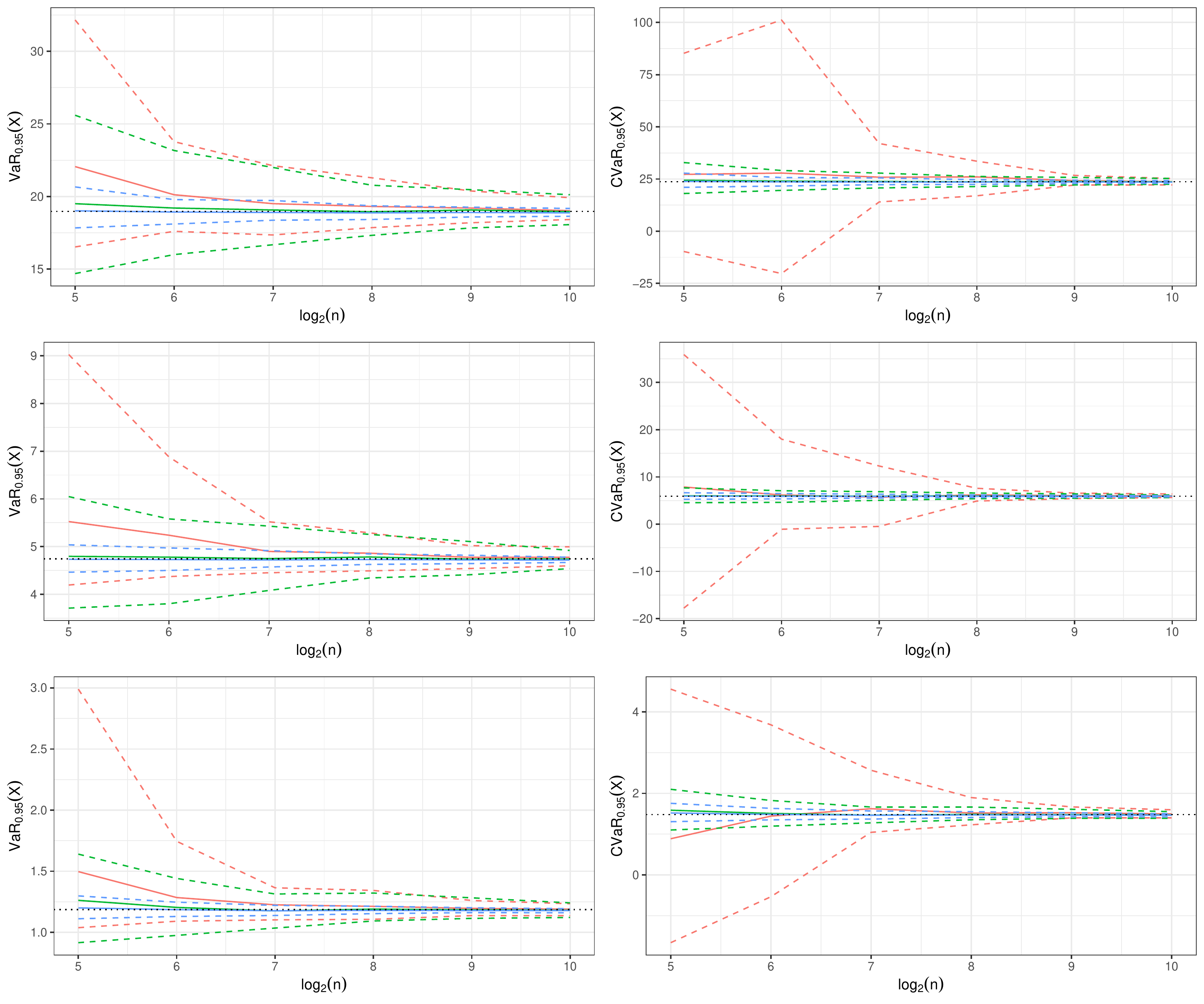}
\end{center}
\caption{Estimates of $\text{VaR}_{0.95} (X)$ (left charts) and $\text{CVaR}_{0.95} (X)$ (right charts), for $X \sim \Gamma(2,\beta)$ with $\beta = 0.25$ (upper charts),  $\beta = 1$ (middle charts) and $\beta = 4$ (lower charts), with 2.5\% and 97.5\% confidence bounds, by MH (red lines), BMH (green lines) and IPBMH (blue lines). Black lines represent the real value of each risk measure.}\label{fig5}
\end{figure}

Let $X \sim \Gamma (\alpha ,  \beta )$.  We considered $\alpha = 2^k,  k = -2, 1,1,2$ and $\beta = 2^j,  j = -2,-1,0,1,2$.  Figures \ref{fig4}-\ref{fig5} show the estimates of $\text{VaR}_{0.95}(X)$ and $\text{CVaR}_{0.95}(X)$ for different values of $\beta$ (0.25,1,4),  for $\alpha = 0.5$ (Fig.  \ref{fig4}) and $\alpha = 2$ (Fig.  \ref{fig5}).

As we can see,  charts from Figure \ref{fig4} are quite similar in shape to those for Figure \ref{fig5}.  As in the previous cases,  MH is the method that offers the worst estimates,  especially when the number of data $n$ is low and with low values of $\beta$.  BMH and IPBMH provide very accurate estimates of both measures,  being IPBMH the most precise method,  regardless of the value of the parameters and the size of the sample.

\section{Application}\label{sec6}

\begin{figure}[htp]
\begin{subfigure}{0.5\textwidth}
\centering
\includegraphics[scale=0.4]{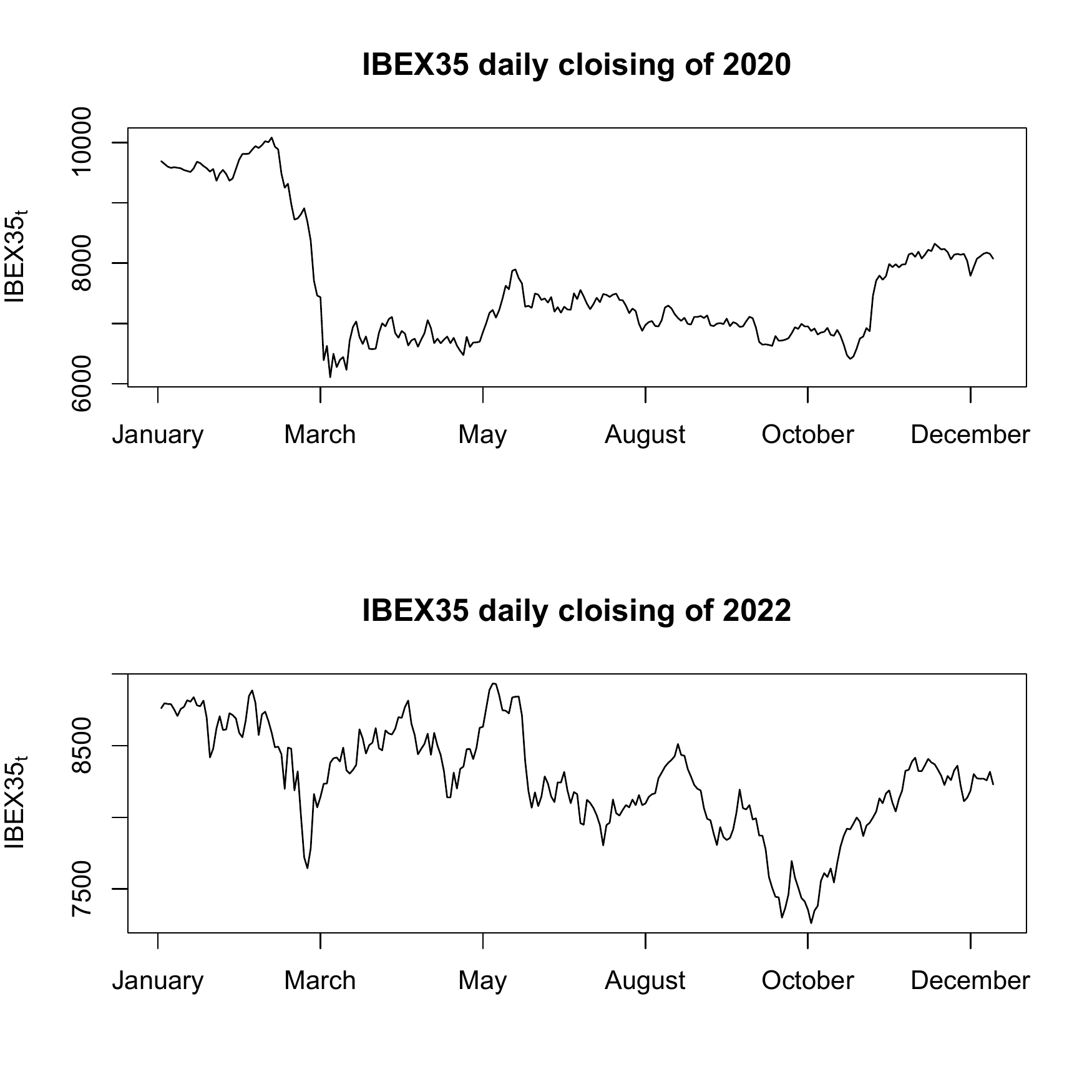}
\end{subfigure} \vspace{0.1cm}
\begin{subfigure}{0.5\textwidth}
\centering
\includegraphics[scale=0.4]{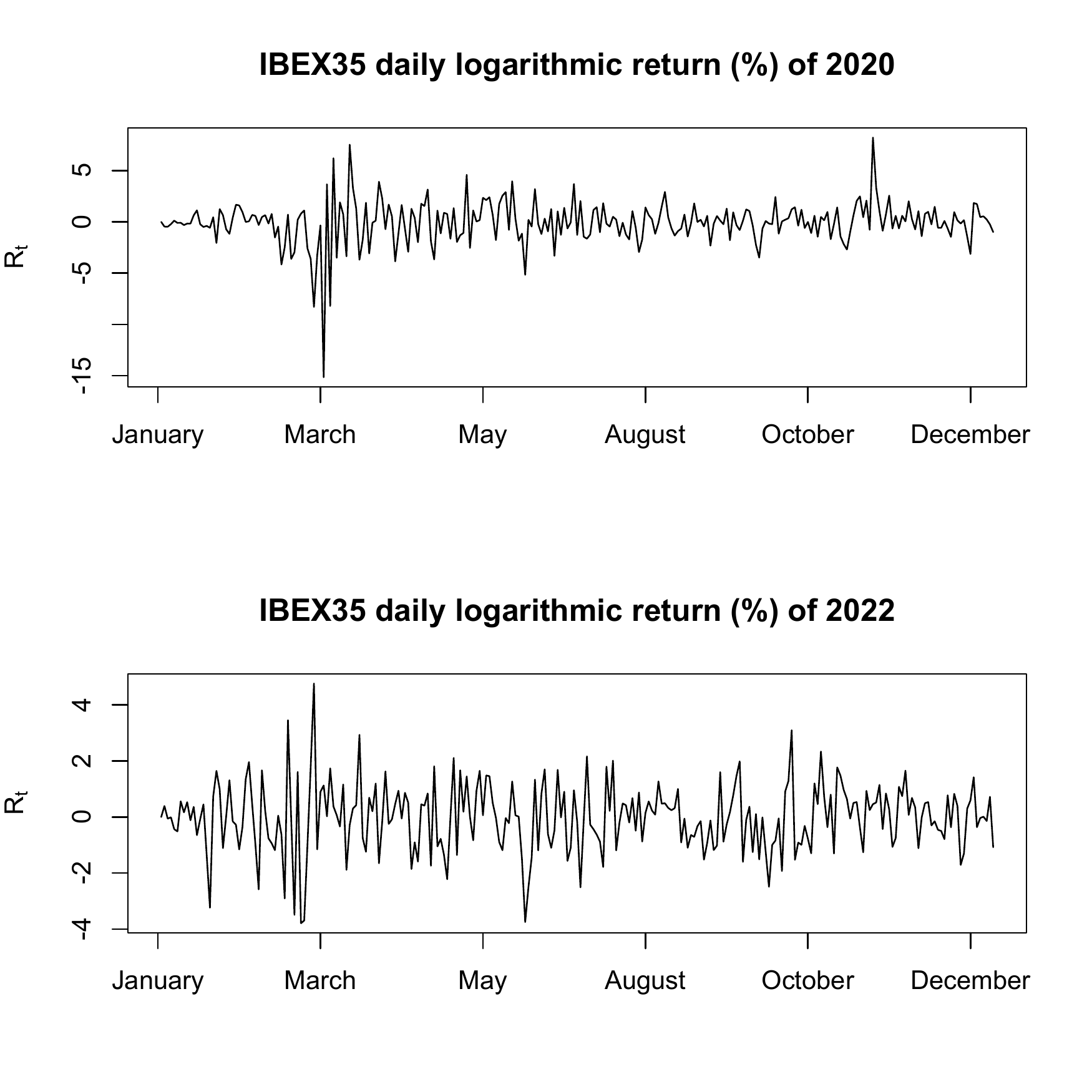}
\end{subfigure}
\caption{IBEX35's daily closing (left charts) and daily logarithmic return (\%) (right charts) for 2020 (upper charts) and 2022 (lower charts). }\label{fig6}
\end{figure}

As we commented previously,  risk measures are essential in Finantial Risk studies.  In Spain,  the most important stock index is Iberia Index or IBEX35.  This index is equivalent to S\&P500 index.  For this study,  the data of IBEX35's daily closing for year 2022 were selected.  Besides,  in order to study the possible consequences of the COVID-19 pandemic about this stock index,  data for year 2020 were taken.  The IBEX35's behaviour is shown in Figure \ref{fig6} (left charts).   

However,  in order to compute the risk measures for financial data,  it is usually employed the daily logarithmic return of stock index.  It can be defined by 

\begin{equation}\label{eq37}
R_t = \ln \left( \dfrac{P_t}{P_{t-1}} \right)
\end{equation} where $P_t$ and $P_{t-1}$ indicate the stock index on day $t$ and $t-1$,  respectively.  

\begin{figure}[htp]
\centering
\begin{subfigure}{0.45\textwidth}
\centering
\includegraphics[scale=0.4]{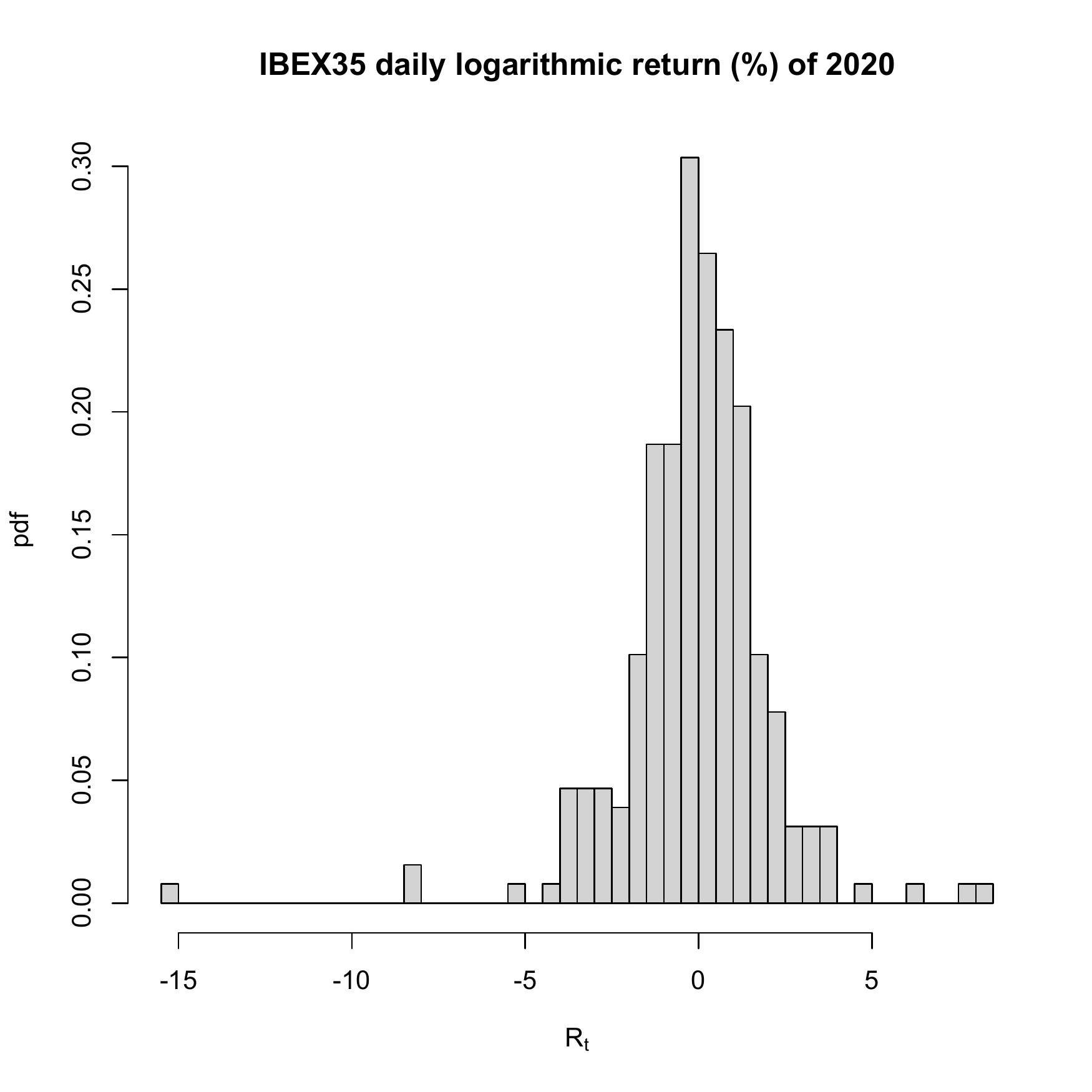}
\end{subfigure} \vspace{0.25cm}
\begin{subfigure}{0.45\textwidth}
\centering
\includegraphics[scale=0.4]{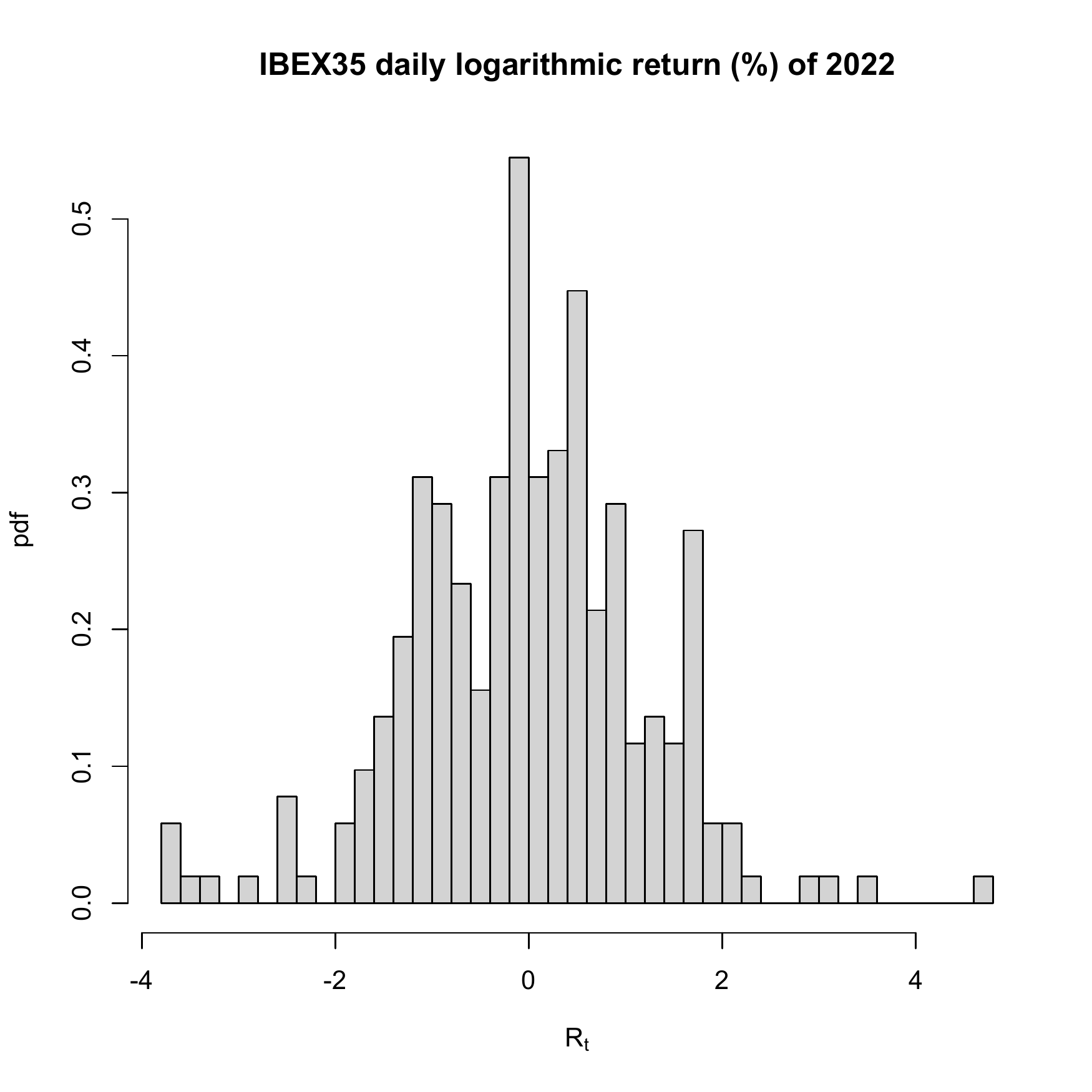}
\end{subfigure}
\caption{Data histogram of IBEX35's daily logarithmic return (\%) of 2020 (left chart) and 2022 (right chart).}\label{fig7}
\end{figure}

Figure \ref{fig6} (right charts) shows the percentage of noise of the one-year daily return  for 2020 and 2022,  respectively.  In both cases,  the majority of these values range between -5.0\% and 5.0\%,  but it should be noted that in March of 2020,  $R_t$ took a value of -15\% due to COVID-19 pandemic.  For both years,  the dataset comprehends a sample of 257 values.  Figure \ref{fig7} shows the behaviour of the data,  and it can be observed that data seem to follow a normal distribution. 

In Finantial Risk,  a quantity of interest is the maximum loss expected in the $95\%$ of the worst cases,  that is,  $\text{VaR}_{0.05}$.  As we have assumed a normal distribution for the dataset,  quantile $\text{VaR}_{0.05}$ is equal to $-\text{VaR}_{0.95}$,  therefore we employ the negative values of $R_t$ for the computations.

To show the advantage of the new proposed method,  IPBMH, a  study was carried out to compare it with MH.  The BMH method can only be used when the distribution of the data is known with certainty,  an assumption that does not generally happen in real situations,  therefore we can not employ BMH method.  

The values that exceed the threshold  $u$ located in the quantile of $p_u = 0.9$ order were considered as extreme values. 

The historical simulation model proposes \cite{Likitratcharoen2021,Likitratcharoen2023} to determine the risk measures employing historical data.  That is,  let $X_t$ be a value of time $t$,  the risk measure value for a time $t$  is computed using $X_1,  ...,  X_{t-1}$ values.  This model needs a large size of data to obtain precise values,  because there are few extreme values. 

\begin{figure}[htp]
\begin{subfigure}{0.5\textwidth}
\centering
\includegraphics[scale=0.4]{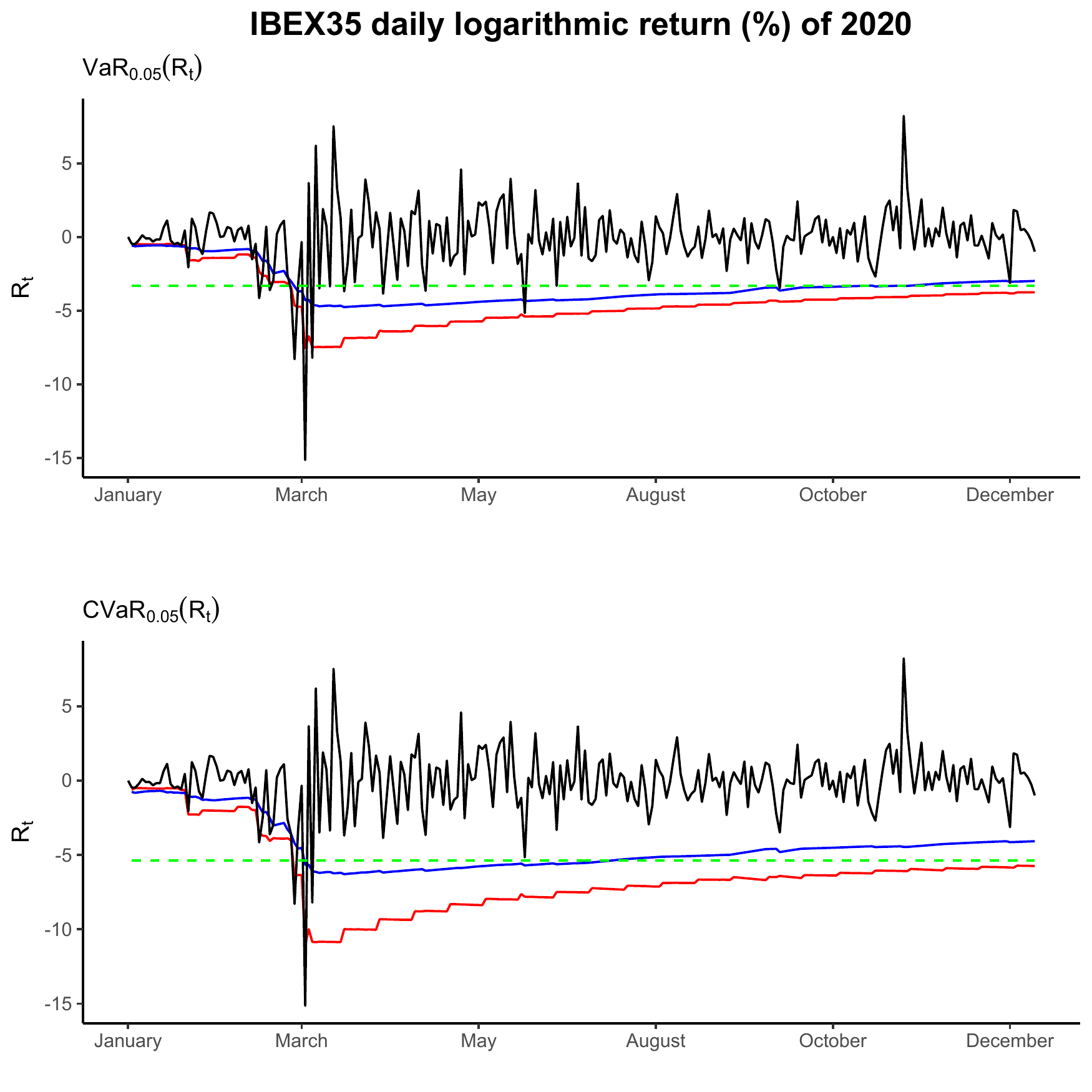}
\end{subfigure} \vspace{0.1cm}
\begin{subfigure}{0.5\textwidth}
\centering
\includegraphics[scale=0.4]{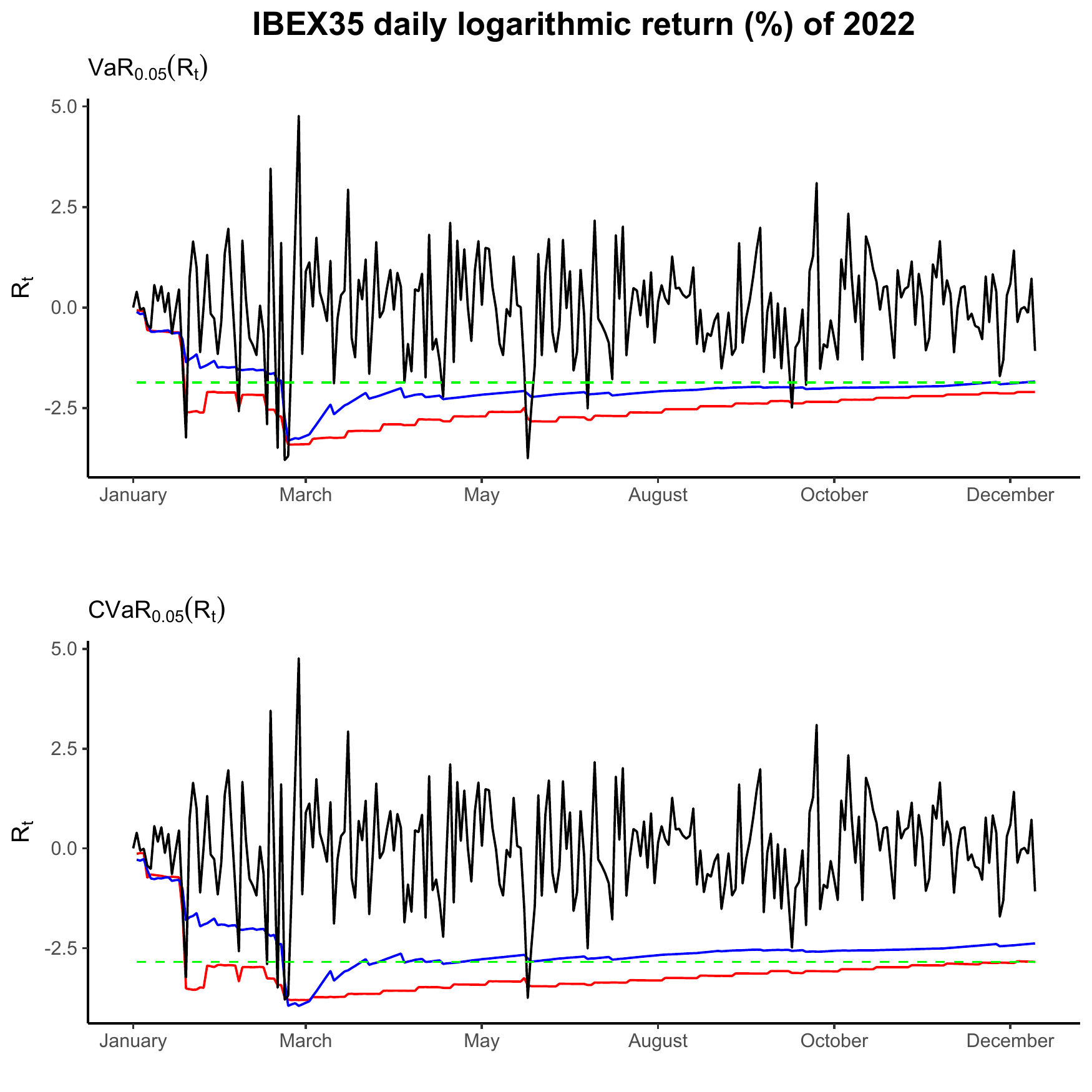}
\end{subfigure}
\caption{Estimates of $\text{VaR}_{0.05} (R_t)$ (upper charts) and $\text{CVaR}_{0.05} (R_t)$ (lower charts) by MH (red lines) and IPBMH (blue lines).  Green dashed lines represent the real value of each risk measure.  Black curves represent the IBEX35 daily logarithmic return (\%).}\label{fig8}
\end{figure}

Figure \ref{fig8} shows the estimates for VaR and CVaR provided by MH (red lines) and IPBMH (blue lines) employing the historical simulation model.  IPBMH provides better estimates than MH,  for any value of the size of the dataset. 

\section{Conclusions}\label{sec7}

\begin{enumerate}
\item We showed analytical expressions to compute VaR and CVaR for most employed distributions in Risk Theory: GPD,  Gamma and Stable distributions (Normal and Cauchy).
\item We detailed two Bayesian methods,  based on MH algorithm,  to estimate VaR and CVaR.  Those methods (MH and BMH) were depeloped for the distributions considered in this work.  BMH consists on considering all the dataset,  not only the observations above the threshold,  to perform estimations.  Therefore,  it is a good strategy when the baseline distribution is known.
\item We proposed a new method wich considers highly informative priors (IPBMH).  This method also employs the whole dataset of observations,  but the tail values have a greater importance than those under the threshold.  It also seizes the relation between the parameters of the baseline distribution and the limit GPD's parameters.  
\item We performed a wide simulation study for the distributions considered,  to compute the accuracy and precision of the estimates for VaR and CVaR given by the three methods.  For any baseline distribution or chosen parameter,  IPBMH provided the most accurate,  less skewed and precise estimates,  especially when the dataset was scarce.  As the sample size increase,  the three methods provide very similar results.
\item BMH and IPBMH present similar behaviour for stable distributions,  regardless of parameters and sample size. 
\item In real situations,  the distribution of the data is generally unknown.  In this case,  we can not apply BMH.  We considered an example with real data and showed how IPBMH is for more appropiate than MH to perform estimation,  with a more stable behaviour and lower dispersion.  
\end{enumerate}

\section*{Declarations}

\begin{itemize}
\item \textbf{Funding}:  This research was part of R\&D\&I Project PID2021-122209OB-C32 funded by 

MCIN/AEI/10.13039/501100011033 and European Regional Development Funds.
%\item Conflict of interest/Competing interests (check journal-specific guidelines for which heading to use)
%\item Ethics approval 
%\item Consent for publication
%\item Availability of data and materials
\item \textbf{Authors' contributions}: These authors contributed equally to this work.
\end{itemize}



\bibliographystyle{elsarticle-num} 
\bibliography{biblio}

\begin{thebibliography}{10}
\expandafter\ifx\csname url\endcsname\relax
  \def\url#1{\texttt{#1}}\fi
\expandafter\ifx\csname urlprefix\endcsname\relax\def\urlprefix{URL }\fi
\expandafter\ifx\csname href\endcsname\relax
  \def\href#1#2{#2} \def\path#1{#1}\fi

\bibitem{Garcia2021}
J.~A. García, M.~M. Pizarro, F.~J. Acero, M.~I. Parra, A bayesian hierarchical
  spatial copula model: An application to extreme temperatures in extremadura
  (spain), Atmosphere 12~(7) (2021).
\newblock \href {https://doi.org/10.3390/atmos12070897}
  {\path{doi:10.3390/atmos12070897}}.

\bibitem{Garcia2018}
J.~Garc{\'\i}a, J.~Mart{\'\i}n, L.~Naranjo, F.~Acero, A bayesian hierarchical
  spatio-temporal model for extreme rainfall in extremadura (spain),
  Hydrological sciences journal 63~(6) (2018) 878--894.
\newblock \href {https://doi.org/10.1080/02626667.2018.1457219}
  {\path{doi:10.1080/02626667.2018.1457219}}.

\bibitem{Longin1998}
F.~M. Longin, Value at risk and extreme values, IFAC Proceedings Volumes
  31~(16) (1998) 45--49.

\bibitem{Chinhamu2015}
K.~Chinhamu, C.-K. Huang, C.-S. Huang, D.~Chikobvu, et~al., Extreme risk,
  value-at-risk and expected shortfall in the gold market, International
  Business \& Economics Research Journal (IBER) 14~(1) (2015) 107--122.
\newblock \href {https://doi.org/10.19030/iber.v14i1.9035}
  {\path{doi:10.19030/iber.v14i1.9035}}.

\bibitem{Embrechts2013}
P.~Embrechts, C.~Kl{\"u}ppelberg, T.~Mikosch, Modelling extremal events: for
  insurance and finance, Vol.~33, Springer Science \& Business Media, Berlin,
  Germany, 2013.

\bibitem{Magnou2017}
G.~Magnou, An application of extreme value theory for measuring financial risk
  in the uruguayan pension fund, Compendium: Cuadernos de Econom{\'\i}a y
  Administraci{\'o}n 4~(7) (2017) 1--19.

\bibitem{ZeaBermudez2010b}
P.~de~Zea~Bermudez, S.~Kotz, Parameter estimation of the generalized pareto
  distribution—part ii, Journal of Statistical Planning and Inference 140~(6)
  (2010) 1374--1388.
\newblock \href {https://doi.org/10.1016/j.jspi.2008.11.020}
  {\path{doi:10.1016/j.jspi.2008.11.020}}.

\bibitem{ZeaBermudez2003}
P.~de~Zea~Bermudez, M.~Turkman, Bayesian approach to parameter estimation of
  the generalized pareto distribution, Test 12~(1) (2003) 259--277.

\bibitem{Diebolt2005}
J.~Diebolt, M.-A. El-Aroui, M.~Garrido, S.~Girard, Quasi-conjugate bayes
  estimates for gpd parameters and application to heavy tails modelling,
  Extremes 8~(1) (2005) 57--78.

\bibitem{Castellanos2007}
M.~E. Castellanos, S.~Cabras, A default bayesian procedure for the generalized
  pareto distribution, Journal of Statistical Planning and Inference 137~(2)
  (2007) 473--483.

\bibitem{Martin2022}
J.~Mart\'in, M.~I. Parra, M.~M. Pizarro, E.~L. Sanju\'an, Baseline methods for
  the parameter estimation of the generalized pareto distribution, Entropy
  24~(2) (2022) 178.
\newblock \href {https://doi.org/10.3390/e24020178}
  {\path{doi:10.3390/e24020178}}.

\bibitem{Martin2020}
J.~Mart\'in, M.~I. Parra, M.~M. Pizarro, E.~L. Sanju\'an, Baseline methods for
  bayesian inference in gumbel distribution, Entropy 22~(11) (2020) 1267.
\newblock \href {https://doi.org/10.3390/e22111267}
  {\path{doi:10.3390/e22111267}}.

\bibitem{Gilli2006}
M.~Gilli, et~al., An application of extreme value theory for measuring
  financial risk, Computational Economics 27~(2) (2006) 207--228.

\bibitem{Bali2007}
T.~G. Bali, A generalized extreme value approach to financial risk measurement,
  Journal of Money, Credit and Banking 39~(7) (2007) 1613--1649.

\bibitem{Trzpiot2010}
G.~Trzpiot, J.~Majewska, Estimation of value at risk: Extreme value and robust
  approaches, Operations Research and Decisions 20~(1) (2010) 131--143.

\bibitem{Park2016}
M.~H. Park, J.~H. Kim, Estimating extreme tail risk measures with generalized
  pareto distribution, Computational Statistics \& Data Analysis 98 (2016)
  91--104.

\bibitem{Van2018}
S.~Van~der Merwe, D.~Steven, M.~Pretorius, Bayesian extreme value analysis of
  stock exchange data, arXiv preprint arXiv:1804.01807 (2018).

\bibitem{El2021}
M.~El~Ghourabi, A.~Nani, I.~Gammoudi, A value-at-risk computation based on
  heavy-tailed distribution for dynamic conditional score models, International
  Journal of Finance \& Economics 26~(2) (2021) 2790--2799.
\newblock \href {https://doi.org/10.1002/ijfe.1934}
  {\path{doi:10.1002/ijfe.1934}}.

\bibitem{Kuang2022}
W.~Kuang, Oil value-at-risk forecasts: A filtered semiparametric approach,
  Journal of Energy Markets 15~(1) (2022).

\bibitem{Korkmaz2018}
M.~{\c{C}}. Korkmaz, E.~Altun, H.~M. Yousof, A.~Z. Afify, S.~Nadarajah, The
  burr x pareto distribution: Properties, applications and var estimation,
  Journal of Risk and Financial Management 11~(1) (2018).
\newblock \href {https://doi.org/10.3390/jrfm11010001}
  {\path{doi:10.3390/jrfm11010001}}.

\bibitem{Nolde2021}
N.~Nolde, C.~Zhou, Extreme value analysis for financial risk management, Annual
  Review of Statistics and Its Application 8 (2021) 217--240.

\bibitem{He2022}
Y.~He, L.~Peng, D.~Zhang, Z.~Zhao, Risk analysis via generalized pareto
  distributions, Journal of Business \& Economic Statistics 40~(2) (2022)
  852--867.
\newblock \href {https://doi.org/10.1080/07350015.2021.1874390}
  {\path{doi:10.1080/07350015.2021.1874390}}.

\bibitem{Faroni2022}
S.~Faroni, O.~Le~Courtois, K.~Ostaszewski, Equivalent risk indicators: Var,
  tce, and beyond, Risks 10~(8) (2022) 142.
\newblock \href {https://doi.org/10.3390/risks10080142}
  {\path{doi:10.3390/risks10080142}}.

\bibitem{McNeil2015}
A.~J. McNeil, R.~Frey, P.~Embrechts, Quantitative Risk Management: Concepts,
  Techniques and Tools Revised edition, no. 10496 in Economics Books, Princeton
  University Press, United Kingdom, 2015.

\bibitem{Balkema1974}
A.~A. Balkema, L.~De~Haan, Residual life time at great age, The Annals of
  probability 2~(5) (1974) 792--804.
\newblock \href {https://doi.org/10.1214/aop/1176996548}
  {\path{doi:10.1214/aop/1176996548}}.

\bibitem{Pickands1975}
J.~Pickands~III, Statistical inference using extreme order statistics, the
  Annals of Statistics (1975) 119--131.

\bibitem{Levy1925}
P.~L{\'e}vy, Calcul des probabilit{\'e}s, Gauthier-Villars, París, 1925.

\bibitem{Nolan2020}
J.~P. Nolan, Univariate stable distributions, Springer, Boston, MA, USA, 2020.

\bibitem{Yamai2002}
Y.~Yamai, T.~Yoshiba, et~al., Comparative analyses of expected shortfall and
  value-at-risk: their estimation error, decomposition, and optimization,
  Monetary and economic studies 20~(1) (2002) 87--121.

\bibitem{Likitratcharoen2021}
D.~Likitratcharoen, N.~Kronprasert, K.~Wiwattanalamphong, C.~Pinmanee, The
  accuracy of risk measurement models on bitcoin market during covid-19
  pandemic, Risks 9~(12) (2021) 222.

\bibitem{Likitratcharoen2023}
D.~Likitratcharoen, P.~Chudasring, C.~Pinmanee, K.~Wiwattanalamphong, The
  efficiency of value-at-risk models during extreme market stress in
  cryptocurrencies, Sustainability 15~(5) (2023) 4395.

\end{thebibliography}


%\begin{thebibliography}{00}
%%% Text of bibliographic item
%
%\bibitem{}
%
%\end{thebibliography}
\end{document}